\documentclass[journal]{IEEEtran}
\usepackage{color}

\usepackage[pdftex]{graphicx}
\graphicspath{{Figs/}}
\usepackage[cmex10]{amsmath}
\usepackage{amssymb}
\usepackage{balance}
\usepackage{algorithmic}
\usepackage{algorithm}
\usepackage{amsfonts}
\usepackage{array}
\usepackage{cite}
\usepackage{mdwmath}
\usepackage[font={footnotesize}]{caption}
\usepackage{subcaption}
\usepackage{stfloats}

\usepackage{pgfplots}
\usepackage{tikz}
\usetikzlibrary{calc}
\usetikzlibrary{arrows,chains,matrix,positioning,scopes}
\makeatletter
\newcommand{\gettikzxy}[3]{%
  \tikz@scan@one@point\pgfutil@firstofone#1\relax
  \edef#2{\the\pgf@x}%
  \edef#3{\the\pgf@y}%
}

\usepackage{colortbl}
\usepackage{amssymb}
\usepackage{pifont}
\newcommand{\cmark}{\ding{51}}%
\makeatother

\begin{document}

\title{Beyond 5G RIS mmWave Systems: Where Communication and Localization Meet}

\author{Jiguang~He,~\IEEEmembership{Member,~IEEE,}
        Fan~Jiang,~\IEEEmembership{Member,~IEEE,}
        Kamran~Keykhosravi,~\IEEEmembership{Member,~IEEE,}
        Joonas~Kokkoniemi,~\IEEEmembership{Member,~IEEE,}
        Henk~Wymeersch,~\IEEEmembership{Senior Member,~IEEE,}
        and~Markku~Juntti,~\IEEEmembership{Fellow,~IEEE}
\thanks{J. He, J. Kokkoniemi, and M. Juntti are with Centre for Wireless Communications, FI-90014, University of Oulu, Finland (E-mail: {firstname.lastname}@oulu.fi).}
\thanks{F. Jiang, K. Keykhosravi, and H. Wymeersch are with Department of Electrical Engineering, Chalmers University of Technology, Gothenburg, Sweden (E-mail: fan.jiang@chalmers.se, kamrank@chalmers.se, henkw@chalmers.se).}
\thanks{This work is supported by Horizon 2020, European Union's Framework Programme for Research and Innovation, under grant agreement no. 871464 (ARIADNE). This work is also partially supported by the Academy of Finland 6Genesis Flagship (grant 318927), the Swedish Research Council (grant no. 2018-03701), and the EU H2020 RISE-6G project.}}

\maketitle

\begin{abstract}
Upcoming beyond fifth generation (5G) communications systems aim at further enhancing  key performance indicators and fully supporting brand new use cases by embracing emerging techniques, e.g., reconfigurable intelligent surface (RIS), integrated communication, localization, and sensing, and mmWave/THz communications. The wireless intelligence empowered by state-of-the-art artificial intelligence techniques has been widely considered at the transceivers, and now the paradigm is deemed to be shifted to the smart control of radio propagation environment by virtue of RISs. In this article, we argue that to harness the full potential of RISs, localization and communication must be tightly coupled. This is in sharp contrast to 5G and earlier generations, where localization was a minor additional service. To support this, we first introduce the fundamentals of RIS mmWave channel modeling, followed by RIS channel state information acquisition and link establishment. Then, we deal with the connection between localization and communications, from a separate and joint perspective. 
\end{abstract}


%
\IEEEpeerreviewmaketitle

\section{Introduction}
\IEEEPARstart{A}{s} the demand on the quality of services (QoSs) keeps rapidly growing,  upcoming beyond fifth generation (B5G) systems are envisioned to meet more stringent requirements, which are beyond those enabled by the ultra-reliable low-latency communication (URLLC), enhanced mobile broadband (eMBB), and massive machine-type communications (mMTC) in the current 5G system. In addition, B5G will also need to support diverse industry vertical applications, e.g., autonomous driving and industry 4.0. Hence, more challenges are brought on physical layer (PHY) transmissions, multi-access, network design, and resource management technologies. A plethora of disruptive techniques, e.g., holographic multiple-input multiple-output (MIMO), reconfigurable intelligent surface (RIS), artificial intelligence (AI) enabled communications and networking, and simultaneous localization and communications (SLAC), are under active investigation to fulfil the targeted QoS requirements for  B5G~\cite{Huang2020,yang2020integrated}.


In order to push data rates, interests towards high-frequency bands, e.g., millimeter wave (mmWave) and even THz, keep increasing~\cite{Hemadeh8207426, Hillger2020}. At these frequencies, the transceivers need to be equipped with a large array of antennas, often called as massive MIMO (mMIMO), to compensate for severe free-space path loss through substantial beamforming gains. Due to the usage of high-resolution analog-to-digital converters (ADCs), a vast amount of power is consumed~\cite{Alkhateeb2014}. When the direct path between transmitter and receiver is blocked, received power drops drastically, adversely affecting QoS. Hence, there is a need for a low-power technology that can overcome link blockages. 
The introduction of RIS to perform analog beamforming towards dedicated users~\cite{2019Basar, Huang2020, Wu2019intelligent} aims to address this need. An RIS is a surface or often a thin covering of a surface with controllable electromagnetic properties. The impedance values of the RIS elements can be controlled so that it realizes an array of phase shifters, thereby modifying the reflection properties and redirecting the impinging wave. Other possible ways to realize  smart surfaces are by metasurfaces, groups of strongly coupled meta-atoms, and holographic surfaces~\cite{DiRenzo9140329}. In addition to reflection, an RIS also has other operational functionalities, e.g., diffraction, refraction, polarization, and absorption, which together makes the intelligent control of wireless propagation channels feasible with enhanced spectral efficiency (SE), energy efficiency (EE), security, and network coverage~\cite{Huang2020}.

\begin{figure*}
     \centering
         \centering
         \includegraphics[width=0.8\textwidth]{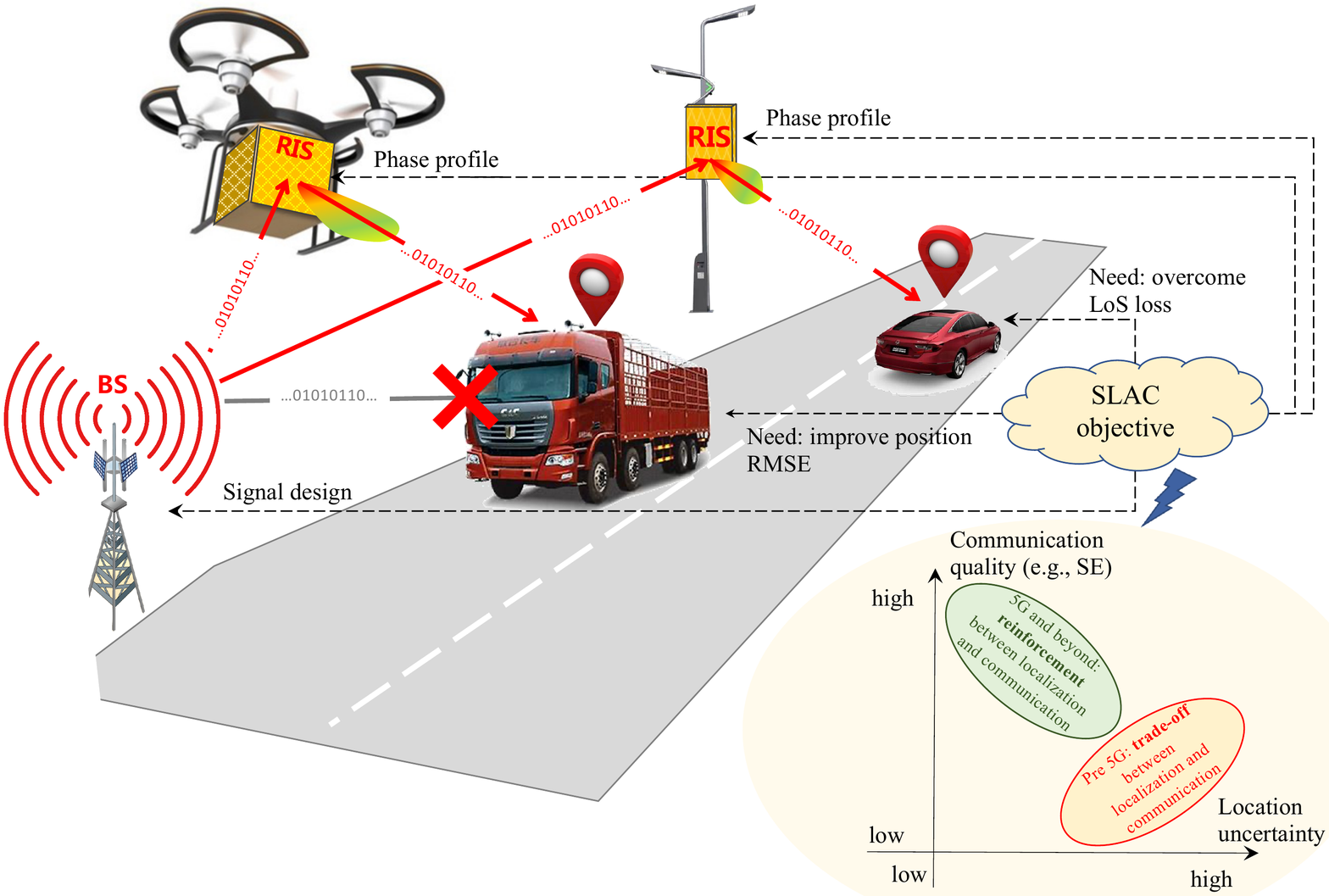}
        \caption{Typical applications for SLAC with RISs. A UAV serves as a mobile location reference and high-gain communication relay, which is configured to meet the MS's SLAC requirements. When the LoS is blocked, the RIS can be reconfigured to provide a strong backup path to the MS. Different from previous generations, localization and communication will reinforce each other, rather than compete for resources.}
        \label{fig:vision_figure}
\end{figure*}

RISs have several characteristics that distinguish them from typical analog arrays. For instance, RISs usually possess a large aperture and a massive number of elements, so the users are likely in the near field. Also, mutual coupling exists among the RIS elements when sub-half-wavelength inter-element spacing is considered~\cite{DiRenzo9140329}. 
To keep the cost down, a passive RIS (with no radio frequency (RF) chains) is usually made of inexpensive components with severe hardware limitations. Hence, when combined with high channel losses and many channel coefficients to be estimated, 
channel estimation (CE) becomes extremely challenging, and may lead to prohibitive training overhead. On the other hand, directional high-gain antennas and a small number of multipath components lead to a sparse geometric channel with lower delay spread and higher coherence bandwidth, which should be exploited during CE. 
Due to the low-rate change of the geometric parameters (user location and orientation), it becomes natural and necessary to harness location information to reduce CE overheads and predict link blockages. Furthermore, an RIS provides a low-cost means to localize users as it is an additional location reference that enables time and angle measurements~\cite{Wymeersch2020}. Hence, the RIS itself provides the localization solution to solve the communication challenges. 
Two exemplary scenarios are unmanned aerial vehicle (UAV) and autonomous driving systems, depicted in Fig.~\ref{fig:vision_figure}, where data transfer between the base station (BS) and mobile station (MS) and radio localization can be simultaneously enabled by RISs. 
While \emph{there have been a large number of magazines and technical papers on the benefits of RIS for communication and localization separately, their intimate connection via SLAC has been largely ignored}. 

In this article, we provide an overview of the key research directions closely related to the realization of SLAC in
RIS mmWave systems: RIS channel modeling and parameter estimation, 
and RIS-aided localization and communications. We first briefly review the salient properties of RIS channels and then describe different approaches to CE  (based on compressive sensing (CS) and data-driven deep learning (DL)). 
Then, in contrast to the majority of the reported works on energy- and spectrum-efficient RIS-aided communications, we specially focus on the connection between localization and communications, as well as SLAC with RISs. 


\section{RIS Channels:  Modeling and Estimation}\label{channel_modeling}

In this section, we give an overview of the high frequency channel modeling of RISs, as well as the related challenges. Then, we show how to exploit the sparse and geometric nature of the channel when establishing links for communications and localization. 



\subsection{RIS mmWave Channel Modeling}\label{channel_model_RIS}
Compared to low frequency systems, the mmWave and THz band channels tend to be simpler, but suffer more from non-line-of-sight (NLoS) propagation losses and hardware challenges (phase noise, nonlinearity)~\cite{Hillger2020}.  
The simplicity of the mmWave channel is due to small number of multipath components, in the addition to the  line-of-sight (LoS) path
~\cite{Hemadeh8207426}. 
With ongoing development  of mmWave and THz, spatio-temporal channel models comprising angle of arrival (AoA), angle of departure (AoD),  time-of-arrival (ToA), delay spread, and distribution of the NLoS paths, the signal processing techniques can be adapted to sparse channel conditions. 
%
%
%
\begin{figure*}[t!]
    \centering
    \includegraphics[width=0.8\linewidth ]{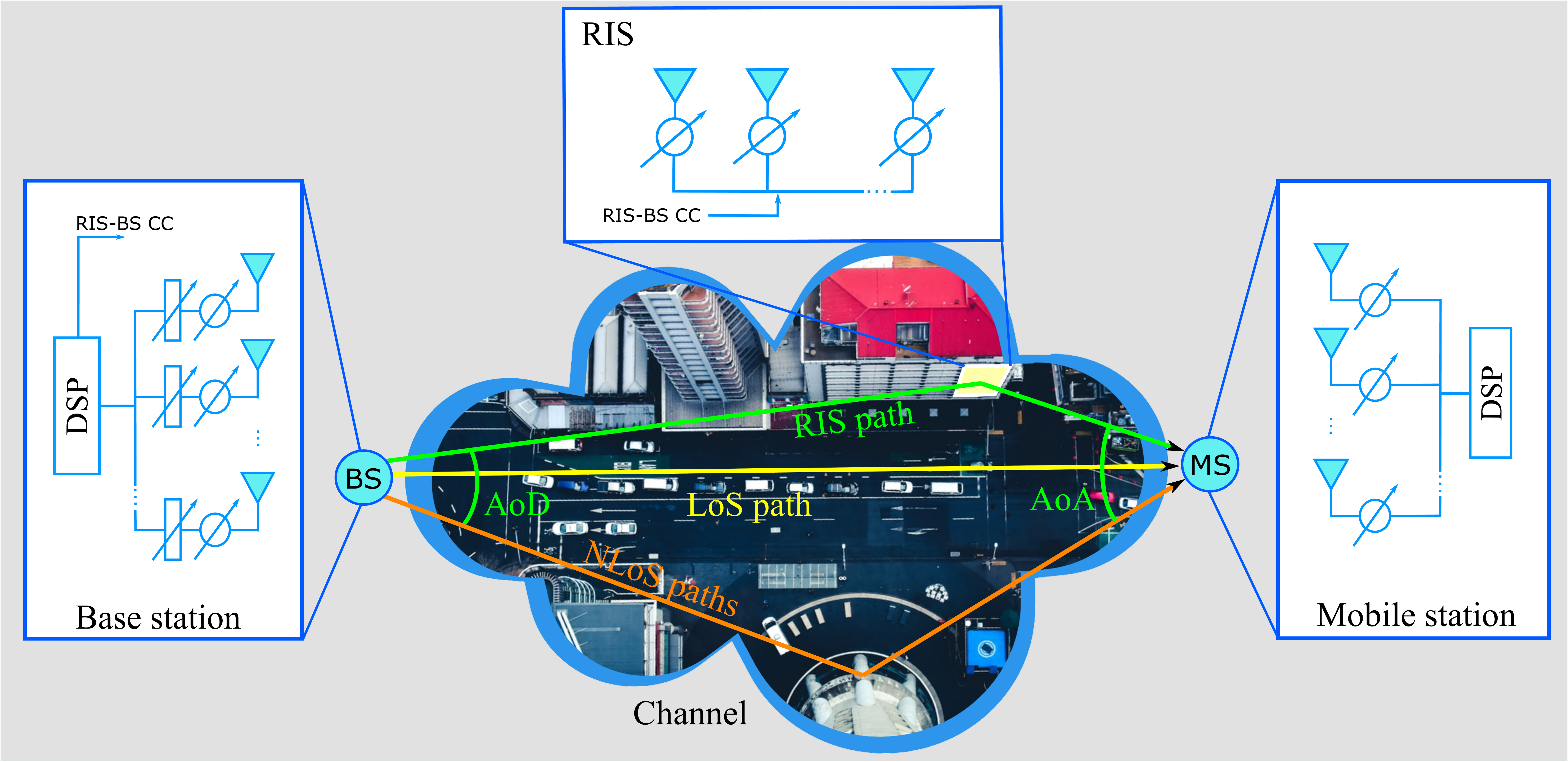}
    \caption{An illustration of an RIS system model with BS and MS antenna arrays and some LoS, NLoS, and RIS-aided channels, as well as AoA and AoD ranges of all possible paths. A control channel (CC) is established between the BS and RIS. }
    \label{fig:Ch}
\end{figure*}
%
%
RISs have mainly gained interest as a means to overcome the large losses at high frequencies by controlling the radio propagation environment
~\cite{2019Basar, Wu2019intelligent, DiRenzo9140329}. Similarly, under optimized phase control, they can provide additional measurements of AoA, AoD, ToA, independent of the uncontrolled multipath, thus enabling or boosting localization and sensing capabilities~\cite{Wymeersch2020}. 
An illustration of a system with an RIS is given in Fig.\ \ref{fig:Ch} with some of the components affecting the BS-MS MIMO channel matrix. These include amplitude and phase control at the BS/MS, the gains of the BS and MS antennas, RIS phase control matrix, and LoS, NLoS and RIS paths. 
The RIS channels are modeled as compound channel between BS-RIS and RIS-MS with the phase control at the RIS. The channel models of the compound link depend on the environment, mobility of users and other network elements, and the frequency band. The phase control at RIS modifies the behavior of the compound link to maximize the channel gains. The RIS phase control depends on various aspects, but predominantly the type of the RIS itself and its limitations (element spacing, element gains, phase shifter resolutions, etc.). As a low-cost RIS will be HW-limited, each of these limitations must be carefully modeled. For instance, realistic models for mutual coupling are needed when RIS elements are spaced less than half a wavelength apart. An additional modeling challenge stems from the large aperture of the RIS, which pushes the far field of the array far away from radiation source, leading to wavefront curvature. 
This near field region can be handled with beamforming algorithms via CE, but may result in decreased channel gains compared to far field operation. 
There are many different types of RISs whose physical properties have impact on the control algorithms. The two most common in literature are phased reflective arrays and metasurface-based RISs \cite{DiRenzo9140329}. The most commonly utilized RIS models in literature are phased reflective arrays as those allow easy way to model and modulate the phase shifts, and subsequently, the beamforming at the RIS. 

In summary, the RIS channel models can be composed by combining the conventional static and mobile channel models with specific RIS hardware models and RIS control algorithms. There are many research challenges ahead to physically and algorithmically understand the behavior of the RISs and channels with RISs, how to estimate the channel, how to steer beams efficiently, and ultimately, how to simultaneously maximize the communication and localization performance.

\subsection{RIS Channel Estimation}\label{channel_estimation}

\begin{figure}[t!]
    \centering
    \includegraphics[width=1\linewidth]{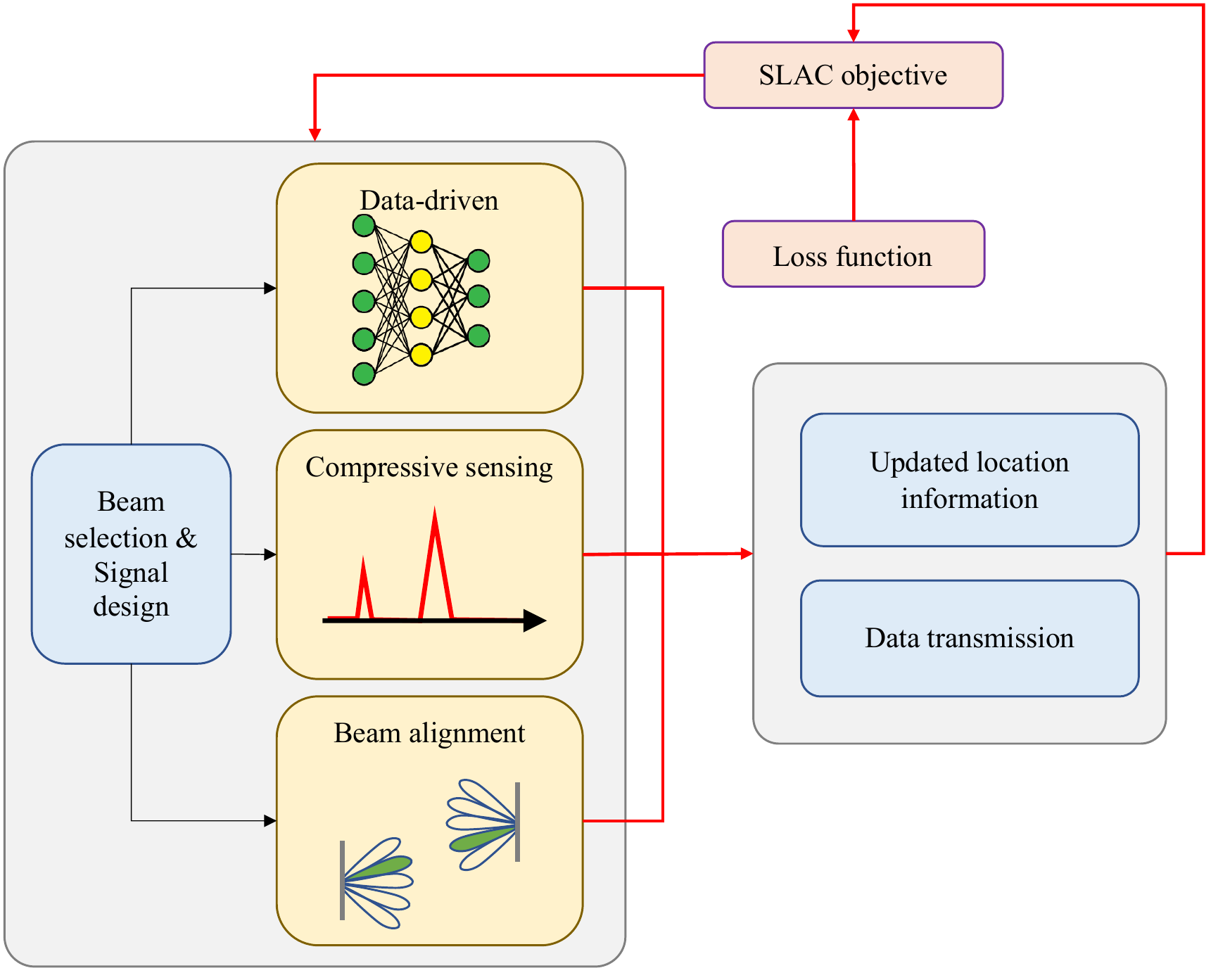}
    \caption{A systematic framework for CS-aided RIS CE, RIS beam alignment, and DNN based data-driven approaches for RIS joint active and passive beamforming, channel parameters extraction, and even MS localization. The beam training matrix design is also of great importance for the above tasks, and can be enhanced by prior location information on MS and environmental objects. As for different SLAC objectives, a variety of loss functions can be considered accordingly. }
    \label{fig:framework}
\end{figure}

 RISs possess several fundamental limitations that affect CE capabilities: (1) a large number of elements but no RF chains; (2) low computational power and small storage space; (3) severe hardware issues such as phase quantization and mutual coupling. All these properties should be taken into consideration when designing and developing efficient yet effective RIS-aided communication and localization protocols.  A systematic framework is sketched in Fig.~\ref{fig:framework}, containing three approaches: pilot-based  approaches relying on CS~\cite{he2020channel}, blind beam alignment, and data-driven approaches. These are described now in detail.
 



%


\subsubsection{RIS Channel Estimation via Compressive Sensing}
In order to achieve the optimal or sub-optimal joint active and passive beamforming, it is essential to get an accurate channel state information (CSI) of all the individual channels or parameters therein. As discussed earlier, there exists inherent sparsity in the mmWave MIMO channels, benefiting from the poor scattering propagation environment, which can be leveraged in the CE algorithm development via advanced CS techniques, e.g., approximate message passing (AMP) and off-the-grid atomic norm minimization (ANM)~\cite{he2020channel}. However, for the purely passive RIS architecture without any connected baseband processing units, CE can only be performed either at the BS via uplink training or the MS via downlink training, due to the lack of observations at the RIS and its limited computation capabilities. When extending to multi-user RIS CE, a shared BS-RIS channel exists among all the users, and this fact should be considered for CE algorithm development. 
In order to simplify the CE, a few active sensors or anchors can be deployed at the RIS, so that CE can be performed at the RIS, by alternations of uplink and downlink training. 

With respect to the beam training matrix, if no prior knowledge about the MS location and environmental objects, e.g., scatters, reflectors, is available, a random construction or selection from part of unitary discrete Fourier transform (DFT) matrix or array response vectors are usually considered. The prior information on MS location and environmental objects can be transformed into the prior knowledge on channel parameters, e.g., AoAs and AoDs. Thus, a better beam training matrix can be designed with higher resolution, compared to the case without any prior information. This in turn brings better performance on RIS CE. On the other hand, the prior information can be used to reduce training overhead by beam selection, resulting in an increase on effective SE. 

\subsubsection{RIS Beam Alignment}
Beam alignment, a one-step direct approach, can intentionally skip the CE process and directly focus on the optimal or suboptimal beam pair selection. It scans and chooses the best pair of beams, one for each terminal, according to a certain criterion, for data transmission. Meanwhile, a coarse MS location can be extracted from the selected beams. The criteria of choosing the best beam pair include maximizing the received power, multiple hypothesis testing, etc. These criteria work well in the high signal-to-noise ratio (SNR) regime and bring a promising performance in terms of beamforming gains and effective SE. Same as RIS CE, the prior location information can also help. 
The weakness lies in that the performance of beam alignment will be determined by the resolution of the predetermined beams (closely related to training overhead) and the criterion of beam selection. When the number of RIS elements is large, its overhead will be inevitably high, which might prevent its usage for RIS-aided mmWave MIMO systems in practice. 
%


\subsubsection{RIS Data-Driven Approaches}
In practice, a huge number of low-cost RIS elements are needed in order to compensate for the severe path loss occurred in the reflection route via RIS. Under such an extreme situation, advanced CS-based CE scheme and standard beam alignment may fail to deliver a satisfactory performance due to either the inevitably high complexity or training overhead. In this regard, data-driven approaches come into effect and may play a pivotal role in CE~\cite{Elbir2020}, joint active and passive beamforming, MS localization, and SLAC (see Fig.~\ref{fig:framework}). The dataset can be collected by fixing the beam training matrices (with reasonable sizes) at all the terminals, resulting in a reasonable training overhead. The labels include the exact (cascaded) channels or parameters therein, joint active and passive beamforming, MS location, or combinations of these. A fully connected DNN or its variant is then trained to map the received signals to the aforementioned labels. Besides, deep unfolding can also be applied with the aid of domain knowledge to mimic the conventional iterative CE algorithms, e.g., iterative reweighted method, projected gradient descent, yielding better estimation performance as well as speeding up the convergence. The training procedure can be done offline with parameter fine-tuning online with a negligible effort. The SLAC loss functions include the combinations of the normalized mean square error (NMSE), SE, and position error bound (PEB). After training, the DL model is used for real-time implementation with the same fixed beam training matrices as in the channel acquisition phase to collect observations. The output of DL model can be used for multiple purposes, as various loss functions can be applied, shown in Fig.~\ref{fig:framework}. 



\subsection{Case Studies}

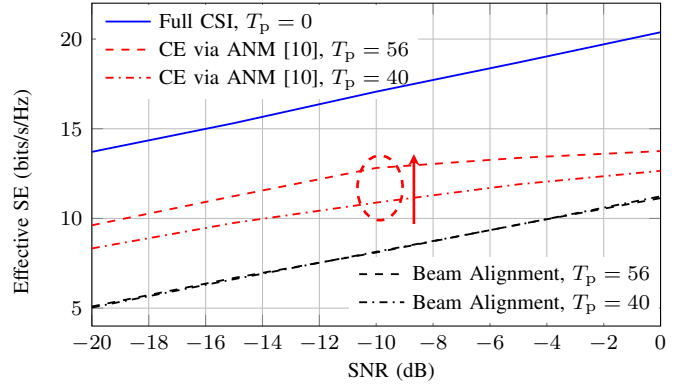
\begin{figure}
\begin{minipage}[!htb]{0.97\linewidth}
    \centering
%
%
\definecolor{mycolor1}{rgb}{0.49020,0.18039,0.56078}%
\definecolor{mycolor2}{rgb}{0.00000,0.45098,0.74118}%
\definecolor{mycolor3}{rgb}{0.47059,0.67059,0.18824}%
\definecolor{mycolor4}{rgb}{1.00000,0.41176,0.16078}%

\begin{tikzpicture}

\begin{axis}[%
width=0.88\textwidth,
height=0.5\textwidth,
at={(0.5,0.5)},
font=\footnotesize,
scale only axis,
xmin=-20,
xmax=0,
ymin=4,
ymax=22,
axis background/.style={fill=white},
xmajorgrids,
ymajorgrids
]
\node (Box2) [draw=none,fill=white] at (axis cs:-5.5,6) {\shortstack[l]{ \ref{BeamAli56}  \footnotesize Beam Alignment, $T_{\mathrm{p}} = 56$ \\
\ref{BeamAli40}  \footnotesize Beam Alignment, $T_{\mathrm{p}} = 40$ }};
\node (Box1) [draw=none,fill=white] at (axis cs:-14.2,19.3) {\shortstack[l]{ \ref{FullCSI} \footnotesize Full CSI, $T_{\mathrm{p}} = 0$ \\
\ref{CEANM56}  \footnotesize CE via ANM~\cite{he2020channel}, $T_{\mathrm{p}} = 56$ \\
\ref{CEANM40}  \footnotesize CE via ANM~\cite{he2020channel}, $T_{\mathrm{p}} = 40$ }};
\addplot [color=blue,line width=.7pt]
  table[row sep=crcr]{%
-20	13.71137375660515\\
-15	15.313403325175726\\
-10	17.067592155161599\\
-5	18.695571204940222\\
0	20.379400453538207\\
};\label{FullCSI}
\addplot [color=red,line width=.7pt,dashed]
  table[row sep=crcr]{%
-20	9.618346896614293\\
-15	11.244978660528444 \\
-10	12.814307872883901\\
-5	13.374517432903421\\
0	13.754524813910148\\
};\label{CEANM56}
\addplot [color=red,line width=.7pt,dashdotted]
  table[row sep=crcr]{%
-20	8.327848736786812\\
-15	9.751570003263401\\
-10	10.88242374641716\\
-5	11.902281928081475\\
0	12.654272029902303\\
};\label{CEANM40}
\addplot [color=black,line width=.7pt,dashed]
  table[row sep=crcr]{%
-20	5.0895\\
-15	6.6830\\
-10	8.1012\\
-5	9.6584\\
0	11.1326\\
};\label{BeamAli56}
\addplot [color=black,line width=.7pt,dashdotted]
  table[row sep=crcr]{%
-20	5.0332\\
-15	6.6129\\
-10	8.1397\\
-5	9.6481\\
0	11.2339\\
};\label{BeamAli40}
\addplot [color=mycolor2, line width=1.5pt, mark size=6.9pt, mark=diamond, mark options={solid, mycolor2}]
  table[row sep=crcr]{%
0	0.303684681653976\\
5	0.116722382605076\\
10	0.0647066384553909\\
15	0.0340290516614914\\
20	0.0203601372241974\\
};
\end{axis}
\node (ylbl) [rotate = 90] at (-0.9,1.8){\footnotesize{Effective SE (bits/s/Hz)}};
\node (xlbl) at (4, -0.6) {\footnotesize{SNR (dB)}};
\draw [color=red, dashed, line width=1pt] (3.85, 1.85) ellipse [x radius=0.3, y radius=0.43];
\draw[-{stealth}, color=red, line width=1pt] (4.3, 1.37) -- (4.3,2.3);
\end{tikzpicture}%
    \caption{Case study 1: CE via compressive sensing (ANM~\cite{he2020channel})  and beam alignment in terms of effective SE, considering 16 BS and 16 MS antennas and 64 RIS elements, all with half-wavelength spacing. Each individual channel is modelled by following Section~\ref{channel_model_RIS}, and the CE is performed at the MS via downlink training. Beam alignment is conducted by creating single-layer beam codebooks at all the terminals. }    \label{fig:Comparison_between_BA_CE}
\end{minipage}
\end{figure}

\begin{figure}
\begin{minipage}[!htb]{0.97\linewidth}
    \centering
%
%
\definecolor{mycolor1}{rgb}{0.49020,0.18039,0.56078}%
\definecolor{mycolor2}{rgb}{0.00000,0.45098,0.74118}%
\definecolor{mycolor3}{rgb}{0.47059,0.67059,0.18824}%
\begin{tikzpicture}

\begin{axis}[%
width=0.88\textwidth,
height=0.5\textwidth,
at={(0.5,0.5)},
font=\footnotesize,
scale only axis,
xmin=0,
xmax=20,
ymode=log,
ymin=0.01,
ymax=1.7489,
axis background/.style={fill=white},
ymajorgrids,
 xmajorgrids,
 ymajorgrids,
yminorgrids,
legend style={at={(0.03,0.03)}, anchor=south west, legend cell align=left, align=left, draw=white!15!black}
]
\addplot [color=red,line width=.7pt,dashed]
  table[row sep=crcr]{
0	0.334585279226303\\
5	0.209359094500542\\
10	0.0786723440885544\\
15	0.0504794605076313\\
20	0.0272031761705875\\
};
\addlegendentry{Deep Unfolding, $T_{\mathrm{p}} = 24$}

\addplot [color=blue,line width=.7pt,dashed]
  table[row sep=crcr]{
0	0.303684681653976\\
5	0.116722382605076\\
10	0.0647066384553909\\
15	0.0340290516614914\\
20	0.0203601372241974\\
};
\addlegendentry{Deep Unfolding, $T_{\mathrm{p}} = 28$}

\addplot [color=black,line width=.7pt]
  table[row sep=crcr]{
0	1.7489\\
5	0.7145\\
10	0.1516\\
15	0.0497\\
20	0.0268\\
};
\addlegendentry{Least Squares, $T_{\mathrm{p}} = 32$}

\end{axis}

\node (ylbl) [rotate = 90] at (-0.9,2.3){\footnotesize{NMSE}};
\node (xlbl) at (4, -0.3) {\footnotesize{SNR (dB)}};
\end{tikzpicture}%
    \caption{Case study 2: CE via data driven approaches (deep unfolding) and least squares (which ignores the channel sparsity) in terms of NMSE, considering 1 BS and 16 MS antennas and 32 RIS elements, all with half-wavelength spacing. Each individual channel is modelled by following Section~\ref{channel_model_RIS}. }    \label{fig:Deep_unfolding}
\end{minipage}
\end{figure}
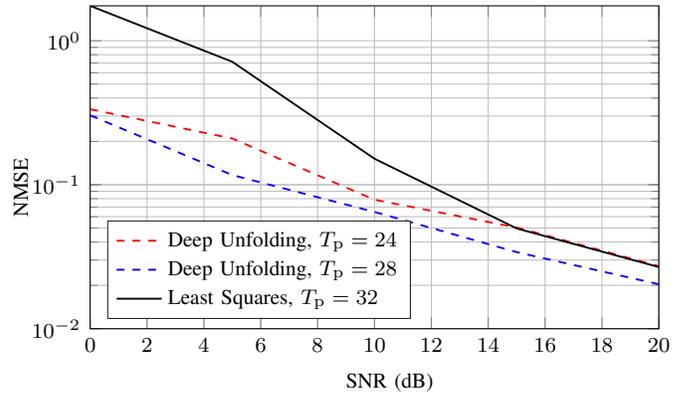

To demonstrate the relative performance of the different approaches, the usefulness of exploiting sparsity, as well as different performance metrics, we have conducted two case studies. The first case study (shown in Fig.~\ref{fig:Comparison_between_BA_CE}) compares RIS channel estimation via compressive sensing  (using ANM~\cite{he2020channel}) with beam alignment, with performance measured in terms of the effective SE (i.e., accounting for the rate loss due to using $T_{\mathrm{p}}$ training signals compared to a frame time of duration $T_{\mathrm{c}}=500$).  
CE via ANM outperforms the standard beam alignment due to its fixed beam resolution. As the effective SE is a function of the training overhead $T_{\mathrm{p}}/T_{\mathrm{c}}$, we observe that the performance does not further improve when we increase the training duration from $T_{\mathrm{p}} = 40$ to $T_{\mathrm{p}} = 56$ in the beam alignment. On the contrary, performance improvement can still be seen in the CE via ANM. 
The second case study (shown in Fig.~\ref{fig:Deep_unfolding}) compares  cascaded channel estimation via data-driven deep unfolding with the standard least squares approach (which ignores channel sparsity), in terms of NMSE. 
Deep unfolding mimics the gradient descent algorithm for a regularized optimization problem, where both channel rank deficiency and noise effect are considered. Deep unfolding based cascaded channel estimation is capable of offering better estimation performance in terms of NMSE compared to the least squares estimation with lower training overhead $T_{\mathrm{p}}/T_{\mathrm{c}}$.



\newcolumntype{P}[1]{>{\centering\arraybackslash}m{#1}}

\begin{table*}[t!]
	\begin{center}
		\centering
		\resizebox{\textwidth}{!} {
			\begin{tabular}{>{\columncolor[gray]{0.9}}P{3cm} P{2cm} P{2cm} |>{\columncolor[gray]{0.9}}P{3cm} P{2cm} P{2cm} } 
				\hline 
				\textbf{Communication  application}&  
				\textbf{Prior location information } & \textbf{Prior map information} & \textbf{Localization and sensing application}  &
				\textbf{Prior location information } & \textbf{Prior map information} 
				\\ \hline
				RIS phase profile optimization & \cmark\cmark\cmark &   \cmark   &  RIS phase profile optimization & \cmark\cmark\cmark & \cmark\cmark
				\\ \hline
				Frame structure optimization & \cmark & -- & User tracking & \cmark\cmark\cmark & \cmark\cmark\cmark  
				\\  \hline
				RIS channel estimation & \cmark \cmark  &   \cmark &   RIS environment mapping &  \cmark & \cmark
				\\  \hline
			   RIS link blockage prediction  & \cmark\cmark  & \cmark\cmark &  RIS object sensing & \cmark & \cmark
				\\ \hline
				 Inter-RIS handover  & \cmark\cmark &  \cmark & Dedicated region (low energy, high secrecy, etc) & \cmark\cmark\cmark & \cmark 
				\\ \hline
			\end{tabular}
		}
		\caption{Overview of SLAC for possible applications in RIS-aided communication. Here, \cmark means that the information has limited impact, \cmark\cmark means that the information is useful, and \cmark\cmark\cmark means that the information provides significant performance benefits. We observe that location and map information can boost RIS-aided communication performance, and should themselves be provided with the aid of RIS.} 
		\label{tab:SLACapps}
	\end{center}
\end{table*}

\section{RIS Simultaneous Localization and Communication} \label{RIS_application}
Many practical applications can be enabled by virtue of RISs. The RIS mmWave MIMO channels contain the geometry information about the environmental objects and the terminals, which inextricably link the two important functionalities, i.e., communications and localization. This also emphasizes the importance of CSI acquisition, already discussed in Section~\ref{channel_estimation}, and the RIS control. An overview of how location and sensing information can be harnessed for communication and localization applications with RIS is provided in Table~\ref{tab:SLACapps}. 
For this reason, we will discuss, in this section, RIS optimization for communications and localization performance separately, as well as their joint performance.  

\subsection{RIS for Communications}

A typical use case of an RIS in wireless communications is to use it as a reflector enabling communications ``around a corner" by using directive beamforming. In mmWave communications, the path loss of the radio wave propagation is very large requiring usually a LoS connection between the transmitter and the receiver. If the direct link is obstructed, a reflected ``LoS"-like link may be created by an RIS. The reflection loss decreases the receive power level, but still it can in many cases be much higher than that caused by random scattering from uncontrolled channel. Due to its large area, an RIS can leverage the near-field effect and create not only directional beams, but also positional beams. Such beams focus energy and provide communication quality in a specific location, rather than a specific direction. 
The RIS operation can be interpreted by at least two ways. One is to see it as a mechanism to \emph{control on the random wireless channel} realization. The classical system design takes the wireless channel as it is so that the waveforms, channel codes, transmit and receive processing etc.\ are the functionalities that an engineer can design. The introduction of an RIS (or multiple RISs) in the channel changes the paradigm so that one can also design the channel to optimize communication-relevant performance metrics.
The other way to interpret the use of an RIS is to interpret it as a \emph{passive relay station}. A true relay station amplifies the received signal directly in the amplify-and-forward (AF) protocol, or decodes and re-transmits it in the decode-and-forward (DF) one. The RIS can operate in some sense similarly to the AF relay without active power amplification or receive noise, because it is a passive element. However, similar order or even better power gains have been reported \cite{9119122-DiRenzo-etal-2020}.


The performance metrics for the communications include the link range or SE, for example. The former can be characterized as the probability of outage of given data rate as a function of the distance between the transmitter and the receiver. The latter is basically the achievable rate or the sum rate in the case of multiple users. Both depend on the signal-to-noise-plus-interference ratio (SINR) at the receiver. The target of the RIS control algorithm is to adjust the phase shifts so that the SINR is maximized by coherent combining of the propagated signal given the instantaneous CSI. 



\subsection{RIS for Localization}
The RIS has been recognized as one of the key enabling technologies toward B5G localization \cite{Wymeersch2020, 6GLocWhitePaper20}. A typical use case on an RIS in localization, is to use it as a location reference point when the LoS connection is obstructed. The reflected LoS path created by the RIS provides an informative measurement of the user's position. Even when LoS is present, the RIS can provide localization improvement, by turning multipath from a foe to a friend, and moreover controlling this multipath to optimize localization performance. In particular, the RIS has great potential to improve the estimation accuracy of the channel parameters (such as AoA, AoD, ToA, phase-of-arrival (PoA), and even Doppler shifts), by optimizing the adjustment of the phase profile or the current distribution in the RIS elements. As a result, the position estimation by utilizing the geometric relationship between the measurements and positions, is significantly improved.


As in communications, RIS operation for localization can be interpreted in several ways. First, an RIS with known position and orientation serves as a \emph{location reference} (similar to a BS) in a global coordinate system. The signal from this reference provides measurements of ToA, AoA, and AoD, which enable more accurate localization. 
A second interpretation is as a \emph{smart object with a unique signature}, providing a distinguishable multipath component, thus aiding localization and environment mapping applications.  

Since an RIS can operate as a reflector, a receiver, and a transmitter. For the most commonly used reflector mode, an RIS can be deployed to smartly forward the wave signals to user terminals or based station, where the multi-path propagation is fully utilized to localize users. 
For the receiver mode, an RIS can be configured with a nearly continuous phase profile, acting as a lens. 
Due to the large size of the RIS, 
wavefront curvature can be harnessed to localize MS and help with system synchronization \cite{AbuKeyKesAleSecWym20}. 
For the transmitter mode, an RIS can 
act similarly to analog transmit beamformer and generate phase-shift keying (PSK) modulated symbols as pilots, hence the channel and position estimation can be performed. 

The performance metrics for localization include root mean-square error (RMSE) of position estimation and the localization coverage. Through Fisher information analysis, the Cram\'{e}r–Rao bound of position can be derived, using as the PEB for localization frameworks. 

\subsection{RIS for Simultaneous Localization and Communications} 

With the development of 5G and beyond, emerging applications have posed \emph{stringent constraints on both communications and localization}. For example, in autonomous driving, a large amount of information needs to be exchanged among vehicles and roadside infrastructure and decimeter-level localization accuracy at the MS side should simultaneously be achieved. In these emerging SLAC applications, the signal and algorithm designs, architecture, and standards, should be further investigated. The RIS can be a promising candidate technology in realizing SLAC applications, since, as we will see, it  shows great potential in achieving high SE communications and high accuracy localization performance.

\begin{figure}
\centering
\input{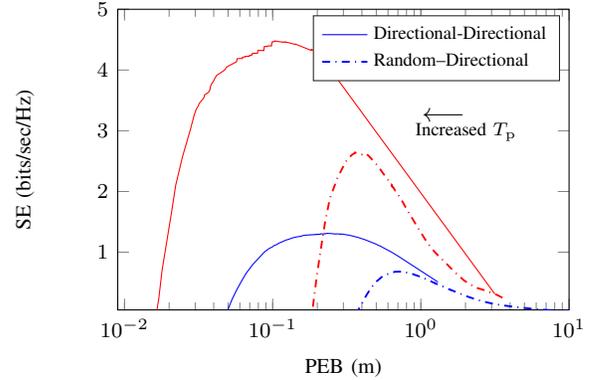}
\caption{ PEB vs. NLoS SE with $T_{\mathrm{p}}$ increasing from right to left, for a SISO system equipped with a single RIS of size $16\times 16$ (blue curves) and $32\times 32$ (red curves). The BS, RIS and the user are located at $(1,1,0)$, $(0,0,0)$, and $(5,5,-5)$, respectively. We consider directional or random RIS profiles for localization and communication. The rest of the parameters can be found in \cite[Table~I]{keykhosravi2020siso}. 
}
\label{fig:SISORIS}
\end{figure}

Distinct from the conventional communications or localization framework, where either the communications metrics (SE, for example) or the localization performance (PEB) are considered, RIS-assisted SLAC requires a proper trade-off between communications and localization. Particularly, the optimal adjustment of the phase profile for the RIS should be designed to meet both the communications and localization requirements. In Fig.\ \ref{fig:SISORIS}, we study the trade--off between communication and localization in a SISO system equipped with a single RIS (as in \cite[Fig.\,1]{keykhosravi2020siso}).
The coherence time is assumed to be $T_{\mathrm{c}} =1000$ symbols, where we use $T_{\mathrm{p}} \geq 3$ symbol slots for transmitting specific pilots and the rest for communication. For each value of $T_{\mathrm{p}}$, the localization and communication performance are calculated in terms of PEB and SE, and illustrated as a curve in Fig.\ \ref{fig:SISORIS}. 
The RIS phase profile sequence can be random (when there is no a priori location information) or directional (when there is a priori location information) during pilot transmission, while it is always directional during communication, as it is based on the estimated channel from the pilots. We observe that such a priori location information can significantly boost both maximal SE (about a factor of 2) and minimal PEB (about a factor of 10). The size of the RIS has a major impact on the performance, where going from 256 elements to 1024 elements improves the maximal SE about 4 times and the minimal PEB about 2--4 times. 
%
%
The figure also clearly illustrates three distinct regimes: (i) communication and localization quality are bad 
when $T_{\mathrm{p}}$ is too small,  as we obtain a poor estimation of the UE position, which leads to directing the signal to an incorrect location and subsequently a low SNR. In this regime, both PEB and SE can be improved by increasing  $T_{\mathrm{p}}$; (ii) good localization performance but low rate occurs when  $T_{\mathrm{p}}$ is too high, since too much of the coherence time is devoted to pilots, leading to excellent channel estimates, but no time to exploit them; (iii) a regime where both positioning and communication have good performance, near the peak of the PEB--SE curves, obtained by moderate values of  $T_{\mathrm{p}}$ (about 5\% overhead for the directional RIS phase profiles and 25\% for the random RIS phase profiles, irrespective of the RIS size).

\section{Outlook}\label{Conclusion}

We have argued that beyond 5G systems enabled by RIS will have to rely on a synergy between communications and localization, since (i) both services rely on the same physical channel; and (ii) one will rely on the other for optimized performance. This leads to the concept of SLAC, which relies on the co-design of communications and localization resources and the flexible trade-off and reinforcement between the two services. To realize SLAC, reliable geometric channel models (including RIS and their hardware effects), as well as low-complexity channel estimation and resource allocation routines must be developed. 
More research and development are required to address the RIS and SLAC technologies towards a wide commercial usage, e.g., in cellular communications, UAV, industry 4.0, autonomous driving, and others.  

\balance

\ifCLASSOPTIONcaptionsoff
  \newpage
\fi

\bibliographystyle{IEEEtran}
\bibliography{IEEEabrv,reference}

\begin{IEEEbiographynophoto}
{Jiguang He} (S'16--M'20) is a postdoctoral researcher at Centre for Wireless Communications (CWC), University of Oulu, Finland. His research interests span millimeter wave MIMO communications, reconfigurable intelligent surfaces for simultaneous localization and  communications.
\end{IEEEbiographynophoto}

\begin{IEEEbiographynophoto}{Fan Jiang} (S'12--M'18) 
is a postdoctoral researcher with Chalmers University of Technology, Sweden.
His research interest includes signal processing for wireless communications, massive multiple-input multiple-output (MIMO) and millimeter wave systems, and localization, tracking, and navigation framework. 
\end{IEEEbiographynophoto}

\begin{IEEEbiographynophoto}{Kamran Keykhosravi}(S'15--M'20)
 is a
 postdoctoral researcher at Chalmers University of Technology, Sweden. His main research interests include  information theory,  reconfigurable  intelligent  surfaces, and radio localization.
\end{IEEEbiographynophoto}

\begin{IEEEbiographynophoto}{Joonas Kokkoniemi} (S'12--M'18) 
is a postdoctoral research fellow at Centre for Wireless Communications, University of Oulu. 
His research interests are in THz band and mmWave channel modeling and communication systems.
\end{IEEEbiographynophoto}

\begin{IEEEbiographynophoto}{Henk Wymeersch}(S'02--M'06--SM'19) received the Ph.D. degree in Electrical Engineering/Applied Sciences in 2005 from Ghent University, Belgium. He 
is currently a Professor in Communication Systems 
at Chalmers University of Technology, Sweden. 
\end{IEEEbiographynophoto}

\begin{IEEEbiographynophoto}{Markku Juntti} (S'93--M'98--SM'04--F'20) is currently a Professor in Centre for Wireless Communications at University of Oulu, Finland. In 1999--2000, he was a Senior Specialist with Nokia Networks. His research interests include signal processing for wireless networks as well as communication and information theory.
\end{IEEEbiographynophoto}

\end{document}